\documentclass{article}

\usepackage{arxiv}

\usepackage[utf8]{inputenc} 
\usepackage[T1]{fontenc}    
\usepackage{hyperref}       
\usepackage{url}            
\usepackage{booktabs}       
\usepackage{amsfonts}       
\usepackage{nicefrac}       
\usepackage{microtype}      
\usepackage{cleveref}       
\usepackage{graphicx}
\usepackage[numbers]{natbib}
\usepackage{doi}
\usepackage{multirow}
\usepackage{array} 

\title{Hibou: A Family of Foundational Vision Transformers for Pathology}
\date{}

\usepackage{authblk}

\author[1]{Dmitry Nechaev\thanks{\texttt{dmitry@hist.ai}}}
\author[1]{Alexey Pchelnikov\thanks{\texttt{alex@hist.ai}}}
\author[1]{Ekaterina Ivanova\thanks{\texttt{kate@hist.ai}}}
\affil[1]{HistAI}

\renewcommand{\headeright}{Hibou Foundational Models for Pathology}
\renewcommand{\undertitle}{}
\renewcommand{\shorttitle}{}

\begin{document}
\maketitle

\begin{abstract}
Pathology, the microscopic examination of diseased tissue, is critical for diagnosing various medical conditions, particularly cancers. Traditional methods are labor-intensive and prone to human error. Digital pathology, which converts glass slides into high-resolution digital images for analysis by computer algorithms, revolutionizes the field by enhancing diagnostic accuracy, consistency, and efficiency through automated image analysis and large-scale data processing. Foundational transformer pretraining is crucial for developing robust, generalizable models as it enables learning from vast amounts of unannotated data.
 
This paper introduces the Hibou family of foundational vision transformers for pathology, leveraging the DINOv2 framework to pretrain two model variants, Hibou-B and Hibou-L, on a proprietary dataset of over 1 million whole slide images (WSIs) representing diverse tissue types and staining techniques. Our pretrained models demonstrate superior performance on both patch-level and slide-level benchmarks, surpassing existing state-of-the-art methods. Notably, Hibou-L achieves the highest average accuracy across multiple benchmark datasets. To support further research and application in the field, we have open-sourced the Hibou models, which can be accessed at \href{https://github.com/HistAI/hibou}{https://github.com/HistAI/hibou}.
\end{abstract}

\section{Introduction}
Pathology is the study of diseased tissue under a microscope, which plays a crucial role in medical diagnosis by allowing pathologists to examine tissue samples to detect abnormalities and disease conditions. It is the gold standard for diagnosing various conditions, particularly cancers, by identifying cellular abnormalities and changes in tissue. Traditional pathology methods involve staining tissue samples and examining them manually under a microscope. While these methods provide detailed insights, they are time-consuming, subject to human error, and heavily reliant on the expertise of the pathologist. Moreover, manual examination limits the scalability and throughput necessary for high-volume clinical settings.

In recent years, there has been a significant shift from traditional pathology to digital pathology, driven by advancements in imaging technology and computational methods. Digital pathology involves scanning conventional glass slides to produce high-resolution digital images, known as whole slide images (WSIs), which can be analyzed using computer algorithms. This transition enhances diagnostic accuracy and efficiency by enabling the use of advanced computational techniques such as machine learning and artificial intelligence (AI). These technologies facilitate automated image analysis, reducing the subjectivity associated with human interpretation and allowing for consistent and reproducible results \citep{Li2020A}.

\section{Related work}
One of the most promising advancements in computational methods for image analysis in digital pathology is the adoption of Vision Transformers (ViTs). ViTs have revolutionized the field of computer vision by achieving state-of-the-art results in various tasks such as image classification, object detection, and segmentation. These models leverage the self-attention mechanism to model long-range dependencies, a fundamental strength over convolutional neural networks (CNNs) which excel at capturing local patterns but struggle with global contexts \citep{Khan2021Transformers}. 

Foundational pretraining techniques for ViTs include supervised learning on large annotated datasets, self-supervised learning where the model is trained using an unlabeled dataset, and transfer learning which involves fine-tuning pre-trained models on new tasks \citep{Han2020A}.
Among these techniques, self-supervised learning stands out as a particularly useful approach since it enables models to learn robust features from unlabeled data, making it valuable in fields like histopathology, where annotated datasets are often limited and costly to produce. By leveraging self-supervised learning, ViTs can be pre-trained on vast amounts of unannotated data, enhancing their ability to generalize and perform well on downstream tasks with limited labeled examples.

Recent works in the field of ViT pretraining for histopathology have predominantly utilized frameworks such as iBot \citep{zhou2022ibot} and DINOv2 \citep{oquab2023dinov2}. The iBot framework is used by the popular open-source model Phikon \citep{Filiot:phikon}. As a more recent and advanced framework, DINOv2 has seen adoption in several notable studies, including Virchow, RudolfV, and Prov-Gigapath, among others \citep{vorontsov2024virchow, dippel2024rudolfv, alfasly2024rotationagnostic, ai2024largescalekaiko, Chen2024uni, Xu2024gigapath}. 

\begin{figure}[htb]
    \centering
    \includegraphics[width=1\linewidth]{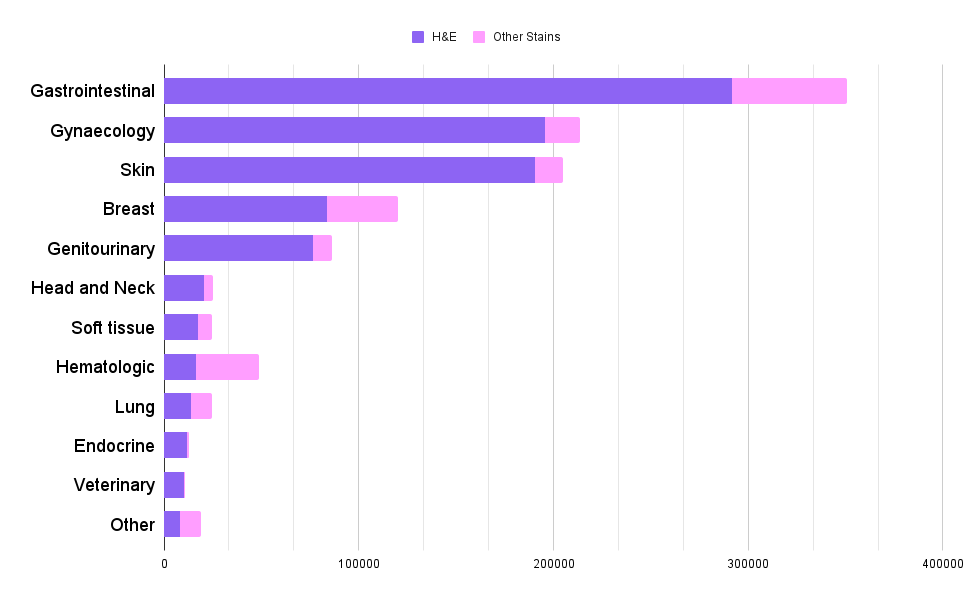}
    \caption{A distribution of tissue types and stains in our dataset}
    \label{fig:datasource}
\end{figure}

In this work, we leverage the DINOv2 framework to pretrain a novel family of vision transformer models, collectively referred to as Hibou. Specifically, we develop two variants: Hibou-B, based on the ViT-B/14 architecture, and Hibou-L, based on the ViT-L/14 architecture. Both models were pretrained on our proprietary histopathology dataset, which comprises over 1 million WSIs representing a diverse array of tissue types and staining techniques (See Figure \ref{fig:datasource} for an overview of the dataset composition).
To promote further research and development, we have made the Hibou-B model publicly available under an Apache 2.0 license. This release is intended to facilitate reproducibility and encourage the application of our pretrained models in various histopathological studies.

\section{Methodology}
\label{sec:methodology}
\subsection{Data}
We trained our foundation models using proprietary data from what we believe to be the most diverse large dataset collected for AI algorithm development. This dataset comprises 936,441 H\&E and 202,464 non-H\&E stained slides sourced from 306,400 unique cases. Our training data includes human tissues from various localizations as well as veterinary biopsies. Additionally, we enriched our dataset with 2,676 cytology slides.

\begin{figure}[htb]
    \centering
    \includegraphics[width=1\linewidth]{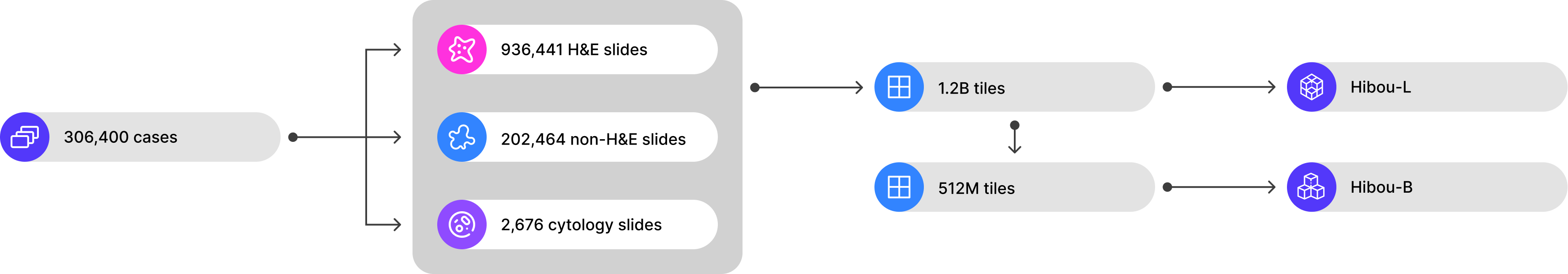}
    \caption{Dataset used for training}
    \label{fig:datasource2}
\end{figure}

To prepare data for training we generate a filtered dataset by splitting WSIs into nonoverlapping patches and filtering out the background patches using Otsu thresholding. In training, we randomly sample tissue patches from the filtered dataset. We use subsets of different sizes depending on the model being trained. For Hibou-L model we use 1.2B clean patches, for Hibou-B we use 512M clean patches. Each unique patch is sampled only once per training.

\subsubsection{Data Augmentations}
DINOv2 uses data augmentations to generate different views of the same image. We use the following set of augmentations in training:
\begin{itemize}
    \item Random angle rotation \citep{alfasly2024rotationagnostic}
    \item Random horizontal and vertical flips
    \item RandStainNA \citep{randstainna}
    \item Color jittering
\end{itemize}

We use RandstainNA in addition to a standard color jittering augmentation as it was shown to improve the performance on WSI-specific downstream tasks \citep{benchmarking:ssl}. We also don't use solarization in line with \citep{dippel2024rudolfv}.

\subsection{Training details}
We use DINOv2 framework \citep{oquab2023dinov2} with registers \citep{darcet2023vitneedreg}. Hibou-B model is trained on 8 A100-80G GPUs with a total batch size of 1024 for 500k iterations. Hibou-L model is trained on 32 A100-40G GPUs with a total batch size of 1024 for 1.175M iterations. Model weights are initialized randomly.

\section{Results}
\label{sec:results}
To evaluate our models we use public datasets and perform evaluation on both patch-level and slide-level tasks. Since our models were trained exclusively on a private dataset it makes the evaluation on public data a fair representation of the ability of our models to generalize to the unseen data.

\subsection{Patch-level benchmarks}
To evaluate a model performance on a patch-level classification task we use a linear probing protocol. We extract features from each image using the pretrained model and then train a linear layer to perform classification. We use SGD as an optimizer and a cosine annealing learning rate. No data augmentations are used in training. For datasets with predefined train-validation-test splits, the official splits are used. In cases where only train-test splits are provided, the training set is randomly partitioned into training and validation subsets. The model checkpoint that achieves the best performance on the validation set is selected, and this checkpoint is then used to evaluate the test set to obtain the final test metrics.

We use the following datasets:
\begin{itemize}
    \item \textbf{CRC-100K}: This publicly available dataset includes 107,180 H\&E-stained images (224×224 pixels) at 20× magnification, obtained from colorectal cancer scans. The images are classified into nine tissue types, representing various components of colorectal tissue, including both healthy and cancerous structures. For our experiments, we utilized only the unnormalized version of the dataset (NCT-CRC-HE-100K-NONORM).
    \item \textbf{MHIST}: Dataset for colorectal polyp classification, consists of 3,152 H\&E-stained images (224×224 pixels). The dataset's primary task is to distinguish between hyperplastic polyps (HP) and sessile serrated adenomas (SSA).
    \item \textbf{PCam}: The PatchCamelyon public dataset comprises 327,680 H\&E-stained images (96×96 pixels). These images are derived from lymph node sections of breast cancer patients and are labeled with binary annotations indicating the presence or absence of metastatic tissue. For testing, we upsampled the images to 224×224 pixels.
    \item \textbf{MSI-CRC}: The dataset comprises 193,312 unique image patches (224×224 pixels, 0.5 µm/px) derived from histological images of colorectal cancer patients in the TCGA cohort. Images are color-normalized using the Macenko method. The dataset is categorized into "MSS" (microsatellite stable) and "MSIMUT" (microsatellite instable or highly mutated) groups. 
    \item \textbf{MSI-STAD}: The dataset comprises 218,578 unique image patches (224×224 pixels, 0.5 µm/px) derived from histological images of gastric cancer patients in the TCGA cohort. Images are color-normalized using the Macenko method. The dataset is categorized into "MSS" (microsatellite stable) and "MSIMUT" (microsatellite instable or highly mutated) groups.
    \item \textbf{TIL-DET}: This dataset consists of 304,097 H\&E images (100×100 pixels, 0.5 µm/px) with or without tumor-infiltrating lymphocytes (TILs) covering 23 different cancer types from the TCGA cohort.
\end{itemize}

\begin{table}[htb]
\caption{Linear probing benchmarks reporting top-1 accuracy. \textsuperscript{*}Metrics for Virchow and RudolfV are derived from the respective papers. \citep{vorontsov2024virchow, dippel2024rudolfv}.}
\centering
\begin{tabular}{>{\raggedright\arraybackslash}p{0.1\linewidth}p{0.08\linewidth}p{0.08\linewidth}p{0.08\linewidth}p{0.08\linewidth}p{0.08\linewidth}p{0.08\linewidth}p{0.08\linewidth}p{0.08\linewidth}}
\toprule
Dataset & Phikon \citep{Filiot:phikon} & Kaiko-B8 \citep{ai2024largescalekaiko} & Virchow \citep{vorontsov2024virchow} & RudolfV \citep{dippel2024rudolfv} & Prov-GigaPath \citep{Xu2024gigapath} & H-optimus-0 \citep{hoptimus0} & Hibou-B & Hibou-L \\
\midrule
{CRC-100K} & 0.917 & 0.949 & 0.968\textsuperscript{*} & \textbf{0.973}\textsuperscript{*} & 0.968 & 0.97 & 0.955 & 0.966 \\
{PCAM} & 0.916 & 0.919 & 0.933\textsuperscript{*} & 0.944\textsuperscript{*} & 0.947 & 0.942 & 0.946 & \textbf{0.953} \\
{MHIST} & 0.791 & 0.832 & 0.834\textsuperscript{*} & 0.821\textsuperscript{*} & 0.839 & \textbf{0.861} & 0.812 & 0.858 \\
{MSI-CRC} & 0.750 & 0.786 & - & 0.755\textsuperscript{*} & 0.771 & 0.767 & 0.779 & \textbf{0.793} \\
{MSI-STAD} & 0.760 & 0.814 & - & 0.788\textsuperscript{*} & 0.784 & 0.797 & 0.797 & \textbf{0.829} \\
{TIL-DET} & 0.944 & 0.945 & - & 0.943\textsuperscript{*} & 0.939 & \textbf{0.948} & 0.942 & 0.942 \\
\midrule
AVG (1-3) & 0.875 & 0.900 & 0.912 & 0.913 & 0.918 & 0.924 & 0.904 & \textbf{0.926} \\
AVG (1-6) & 0.846 & 0.874 & - & 0.871 & 0.875 & 0.881 & 0.872 & \textbf{0.890} \\
\bottomrule
\end{tabular}
\label{tab:linearprobe}
\end{table}

Hibou-L achieves the highest average accuracy across all six datasets, as indicated in Table~\ref{tab:linearprobe}, setting new state-of-the-art performance. The consistent performance across multiple datasets demonstrates the robustness of Hibou-L in handling various histopathological tasks. This robustness is critical for practical applications in clinical settings, where variability in tissue samples can be significant.

\subsection{Slide-level benchmarks}
\label{subsec:slide-level}
We evaluate our model on a classification task using publicly available datasets hosted on The Cancer Genome Atlas (TCGA). We use a weakly supervised approach where each slide is divided into nonoverlapping foreground patches and each sequence of patches corresponding to a single slide is assigned a single label.
For feature extraction, we utilize a pretrained model to generate features for each patch. Then we use a pooling model based on the attention mechanism to aggregate these feature sequences and perform classification. During the training process, only the parameters of the pooling model are updated, while the parameters of the feature extractor remain frozen. We use the AdamW \citep{loshchilov2019decoupledadamw} optimizer for training and do not apply any data augmentations.

The evaluation is conducted on the following datasets:
\begin{itemize}
    \item \textbf{BRCA}: A TCGA-BRCA project, containing 963 WSIs that are labeled: infiltrating duct carcinoma (767 WSIs) or lobular carcinoma (196 WSIs).
    \item \textbf{NSCLC}: A combination of TCGA-LUAD and TCGA-LUSC projects, containing 973 WSIs that are labeled: squamous cell carcinoma (520 WSIs) or adenocarcinoma (453 WSIs).
    \item \textbf{RCC}: A combination of TCGA-KIRC, TCGA-KIRP, and TCGA-KICH projects, containing 927 WSIs that are labeled: renal cell carcinoma (113 WSIs), clear cell adenocarcinoma (523 WSIs), papillary adenocarcinoma (291 WSIs).
\end{itemize}

Each dataset is divided into training, validation, and test subsets following an 80:10:10 ratio. The model is trained using the training subset, and its performance is monitored on the validation subset. We select the checkpoint with the highest validation performance and use this model to evaluate the test subset.

\begin{table}[htb]
	\caption{AUC, WSI subtyping benchmarks, test subset}
	\centering
	\begin{tabular}{llll}
		\toprule
		Dataset & Prov-GigaPath\citep{Xu2024gigapath} & Hibou-B & Hibou-L \\
		\midrule
		BRCA & 0.918 & 0.929 & \textbf{0.946} \\
		NSCLC & 0.967 & 0.952 & \textbf{0.969} \\
		RCC & 0.987 & 0.993 & \textbf{0.996} \\
		\bottomrule
	\end{tabular}
	\label{tab:slideprobe}
\end{table}

Hibou-L achieves the highest AUC across all three datasets, as shown in Table~\ref{tab:slideprobe}, while Hibou-B surpasses Prov-GigaPath\footnote{For our testing, we utilized only the tile-level feature extractor from GigaPath in conjunction with our pooling model, omitting the use of the slide-level feature extractor from GigaPath. This was done to compare tile-level models in the same setting.} in two out of three benchmarks despite having 13 times fewer parameters. This achievement underscores the advanced capabilities of the Hibou models in generating high-quality patch-level features that contribute to accurate slide-level predictions. The efficiency of Hibou-B, in particular, highlights its potential for practical applications where computational resources may be limited, yet high performance is still required. The consistent top performance of Hibou-L across diverse datasets further demonstrates its robustness and adaptability in handling various histopathological classification tasks.

\subsection{Segmentation benchmarks}

To evaluate the performance of the Hibou model on the segmentation task we employed a CellViT \citep{hörst2023cellvitvisiontransformersprecise} framework and trained a segmentation model on a PanNuke dataset \citep{gamper2019pannuke, gamper2020pannuke}. PanNuke is an extensive, open-source pan-cancer histology dataset for nuclei instance segmentation and classification. We follow the training protocols described in \citep{hörst2023cellvitvisiontransformersprecise}, we train the model 3 times on different PanNuke splits and report averaged metrics. For more information on metrics and protocols check the original CellViT paper.

\begin{table}[htb]
\caption{Average PQ across the three PanNuke splits for each nuclear category. Metrics for CellViT\(_{256}\) and CellViT-SAM-H are from \citep{hörst2023cellvitvisiontransformersprecise}.}
\centering
\begin{tabular}{lccccc}
\toprule
\textbf{Model} & \textbf{Neoplastic} & \textbf{Epithelial} & \textbf{Inflammatory} & \textbf{Connective} & \textbf{Dead} \\
\midrule
CellViT\(_{256}\) & 0.567 & 0.559 & 0.405 & 0.405 & 0.144 \\
 CellViT-SAM-H & {0.581} & {0.583} & 0.417 & {0.423} & 0.149 \\
CellViT-Hibou-L & \textbf{0.582} & \textbf{0.591} & \textbf{0.426} & \textbf{0.425} & \textbf{0.185}  \\
\bottomrule
\end{tabular}
\label{tab:cellvit_performance}
\end{table}

\begin{table}[htb]
\caption{Average Precision, Recall, and F\textsubscript{1} across the three PanNuke splits for each nuclear category. Metrics for CellViT\(_{256}\) and CellViT-SAM-H are from \citep{hörst2023cellvitvisiontransformersprecise}.}
\centering
\scriptsize
\begin{tabular}{lcccccccccccccccc}
\toprule
\textbf{Model} & \multicolumn{3}{c}{\textbf{Neoplastic}} & \multicolumn{3}{c}{\textbf{Epithelial}} & \multicolumn{3}{c}{\textbf{Inflammatory}} & \multicolumn{3}{c}{\textbf{Connective}} & \multicolumn{3}{c}{\textbf{Dead}} \\
 & \textit{P} & \textit{R} & \textit{F\textsubscript{1}} & \textit{P} & \textit{R} & \textit{F\textsubscript{1}} & \textit{P} & \textit{R} & \textit{F\textsubscript{1}} & \textit{P} & \textit{R} & \textit{F\textsubscript{1}} & \textit{P} & \textit{R} & \textit{F\textsubscript{1}} \\
\midrule
CellViT\(_{256}\) & 0.69 & 0.70 & 0.69 & 0.68 & 0.71 & 0.70 & 0.59 & \textbf{0.58} & 0.58 & 0.53 & 0.51 & 0.52 & 0.39 & \textbf{0.35} & 0.37 \\
CellViT-SAM-H & \textbf{0.72} & 0.69 & {0.71} & 0.72 & {0.73} & {0.73} & 0.59 & {0.57} & 0.58 & {0.55} & {0.52} & {0.53} & {0.43} & {0.32} & {0.36} \\
CellViT-Hibou-L & \textbf{0.72} & \textbf{0.72} & \textbf{0.72} & \textbf{0.76} & \textbf{0.76} & \textbf{0.76} & \textbf{0.63} & \textbf{0.58} & \textbf{0.60} & \textbf{0.58} & \textbf{0.53} & \textbf{0.55} & \textbf{0.51} & \textbf{0.35} & \textbf{0.41} \\
\bottomrule
\end{tabular}
\label{tab:cellvit_metrics}
\end{table}

As demonstrated in Table~\ref{tab:cellvit_performance} and Table~\ref{tab:cellvit_metrics} the CellViT-Hibou-L model, which uses the Hibou-L architecture as its backbone, consistently outperforms both the smaller ViT\(_{256}\) model, pretrained on pathological data \citep{cell256}, and a larger SAM-H \citep{kirillov2023segany} model, developed specifically for segmentation tasks but trained on natural images.

\section{Discussion and Future Work}

In this study, we introduced the Hibou family of vision transformer models, leveraging the DINOv2 framework for self-supervised pretraining on histopathology data. Despite the promising results, the Hibou-L model has only been trained on approximately one-sixth of our full dataset. We anticipate that further training on more data will enhance the model's performance metrics, as additional data often leads to improved generalization and robustness in Vision Transformers.

Future work will focus on expanding our evaluation benchmarks to include additional subtyping tasks, which are critical for comprehensive histopathological analysis. Furthermore, we plan to investigate slide-level pretraining as this approach has the potential to improve the performance on WSI downstream tasks.
Another promising direction for future research involves utilizing Hibou models as vision encoders in Large Vision-Language Models (LVLMs). These models integrate visual and textual data, enabling sophisticated interactions with histopathological slides. For instance, an LVLM could allow pathologists to query the model in natural language about specific features or abnormalities observed in a slide, receive detailed explanations, and even generate descriptive reports. This interactive capability could enhance diagnostic accuracy, streamline workflows, and facilitate a more intuitive and comprehensive analysis of histopathological data.

We have open-sourced the Hibou-B model to support further research and development in the community. The model is available for a wide range of applications, including commercial use, and can be accessed at \href{https://github.com/HistAI/hibou}{https://github.com/HistAI/hibou}. We encourage researchers and practitioners to build upon our work, contributing to the advancement of AI in histopathology.

\section{Acknowledgments}
We gratefully acknowledge The Cancer Genome Atlas (TCGA) Research Network for providing the publicly available datasets used in this study. The data utilized in this research was obtained from the TCGA data portal at \href{https://portal.gdc.cancer.gov/}{https://portal.gdc.cancer.gov/}.

\bibliographystyle{unsrtnat}
\bibliography{references}
\end{document}